\documentclass[twoside]{ae100prg}
\usepackage{graphicx}
\usepackage[breaklinks]{hyperref}
\usepackage{booktabs}

\begin{document}
\title{Finsler spacetimes and gravity}


\author{Christian Pfeifer$^1$, Mattias Wohlfarth$^1$}

\address{$^1$ II. Institut f\"ur Theoretische Physik und Zentrum f\"ur Mathematische Physik, Universit\"at Hamburg, Luruper Chaussee 149, 22761 Hamburg, Germany}

\email{christian.pfeifer@desy.de}

\begin{abstract}
We consider the geometry of spacetime based on a non-metric, Finslerian, length measure, which, in terms of physics, represents a generalized clock. Our definition of Finsler spacetimes ensure a well defined notion of causality, a precise description of observers and a geometric background for field theories. Moreover we present our Finsler geometric extension of the Einstein equations, which determine the geometry of Finsler spacetimes dynamically.
\end{abstract}

\section{Introduction}
For a hundred years Lorentzian manifolds serve as geometric background for physics. Equipped with the standard model of particle physics this led to the explanation of a huge amount of observations. However, on this basis  we have to conclude that 96\% of the universe are unknown; called dark matter and dark energy \cite{Spergel-07}. Today most explanation attempts for this fact come from modifications of the standard model of particle physics; but possibly a well controlled extension of the geometric background  for physics is able to shed light on the dark universe.

Here we present Finsler spacetimes which are capable to serve as generalized geometric background for physics providing: 
\begin{itemize}
	\item a precise well-defined notion of causality,
	\item a notion of observers and their measurements,
	\item a geometric background for field theories,
	\item and gravitational dynamics consistent with general relativity.
\end{itemize}
Further details beyond this invitation can be found in our articles \cite{Pfeifer:2011tk,Pfeifer:2011xi}.

\section{Finsler geometry}
One of the fundamental measurements in physics is the measurement of time. Its theoretical description is given by EinsteinÕs clock postulate: The time that passes for an observer between two events is given by the length of the observers worldline connecting the events. In case the geometry of spacetime is fundamentally determined by a metric this length is given by
\begin{equation}
S[x]=\int d\tau \sqrt{g_{ab}(x)\dot x^a\dot x^b}\,.
\end{equation}
The key idea for Finsler spacetimes is a more general description of the measurement of time which still realizes the weak equivalence principle: 
\begin{equation}
S[x]=\int d\tau F(x,\dot x)\,.
\end{equation}
It is based on a one-homogeneous function $F$ on the tangent bundle which determines the geometry of spacetime. This so called Finsler geometry is a well known mathematical framework which extends Riemannian metric geometry \cite{Bao}. However this standard  Finsler geometry brakes down as soon as $F$ has a non-trivial null-structure $N_x=\{y\in T_xM|F(x,y)=0\}$. For generalizations of the Lorentzian metric length measures we introduce our definition of Finsler spacetimes which ensure the existence of a precise notion of causality and the existence of a well-defined geometry. 

\section{Causality}
The description of Finsler spacetimes requires the tangent bundle TM of the spacetime manifold M. We consider the tangent bundle in manifold induced coordinates $(x,y)=Z\in TM, Z=y^a\partial_a{}_{|x}$ and its tangent spaces $T_{(x,y)}TM$ in the coordinate basis $\{\partial_a=\frac{\partial}{\partial x^a},\bar \partial_a=\frac{\partial}{\partial y^a}\}$.

A Finsler spacetime $(M,L,F)$ is a smooth manifold $M$ equipped with a continuous function $L:TM\mapsto\mathbb{R}$ such that
\begin{itemize}
	\item $L$ is smooth on the tangent bundle without the zero section $TM\setminus\{0\}$,
	\item $L$ is reversible $|L(x,-y)|=|L(x,y)|$,
	\item $L$ is positively homogeneous of degree $r\geq2$: $L(x,\lambda y)=\lambda^rL(x,y)$,
	\item $g^L{}_{ab}=\frac{1}{2}\bar\partial_a\bar\partial_bL$ is non-degenerate on $TM\setminus A$, $A\subset TM$ measure zero,
	\item $\forall x\in M$ there exists a non-empty closed connected set $S_x\in T_xM$ where: $|L(x,y)|=1$ and $sign( g^L{}_{ab})=(\epsilon,-\epsilon,-\epsilon,-\epsilon)$ with $\epsilon=\frac{|L(x,y)|}{L(x,y)}$.
\end{itemize}
The Finsler function $F$, which defines the geometric clock is a derived object and defined as $F=|L|^{\frac{1}{r}}$; the Finsler metric is $g^F{}_{ab}=\frac{1}{2}\bar\partial_a\bar\partial_bF^2$.

\begin{figure}[h!]
  \begin{center}
  \includegraphics[width=0.4\textwidth]{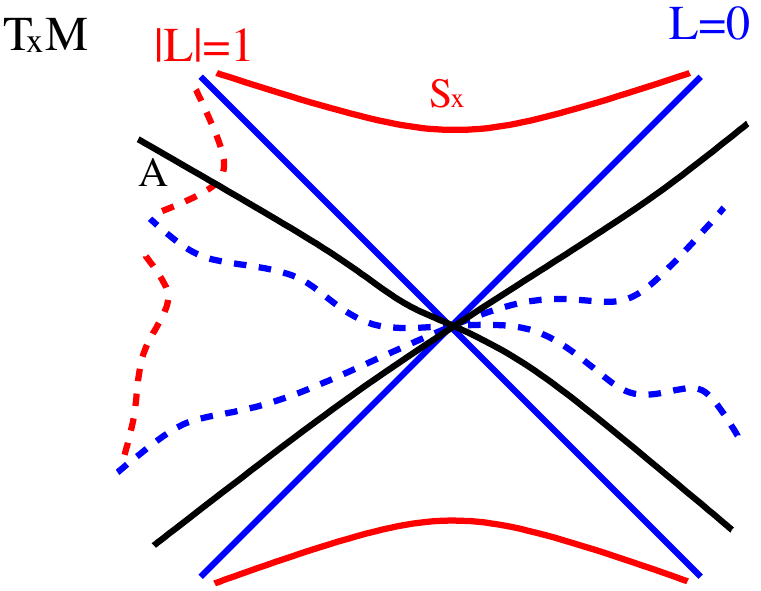}
  \end{center}
  \caption{\label{fig:fst}Causal structure of Finsler spacetime.}
\end{figure}

Our definition of Finsler spacetimes guarantees a causal structure in each tangent space: $S_x$ is the shell of unit timelike vectors which defines a cone of timelike directions with null boundary, as displayed in figure \ref{fig:fst}.

The geometry of Finsler spacetimes is solely derived from derivatives of $L$ in terms of the unique Cartan non-linear connection coefficients: $N^a{}_b=\frac{1}{4}\bar\partial_b(g^{Laq}(y^m\partial_m\bar\partial_qL-\partial_qL))$. The connection between our definition of Finsler spacetimes and standard Finsler geometry is given by the following theorem: \textit{Wherever L and F are both differentiable they encode the same geometry, i.e. $N[L]=N[F^2]$.}

\section{Observers and Matter fields}
\begin{figure}[h]
  \begin{center}
  \includegraphics[width=0.5\textwidth]{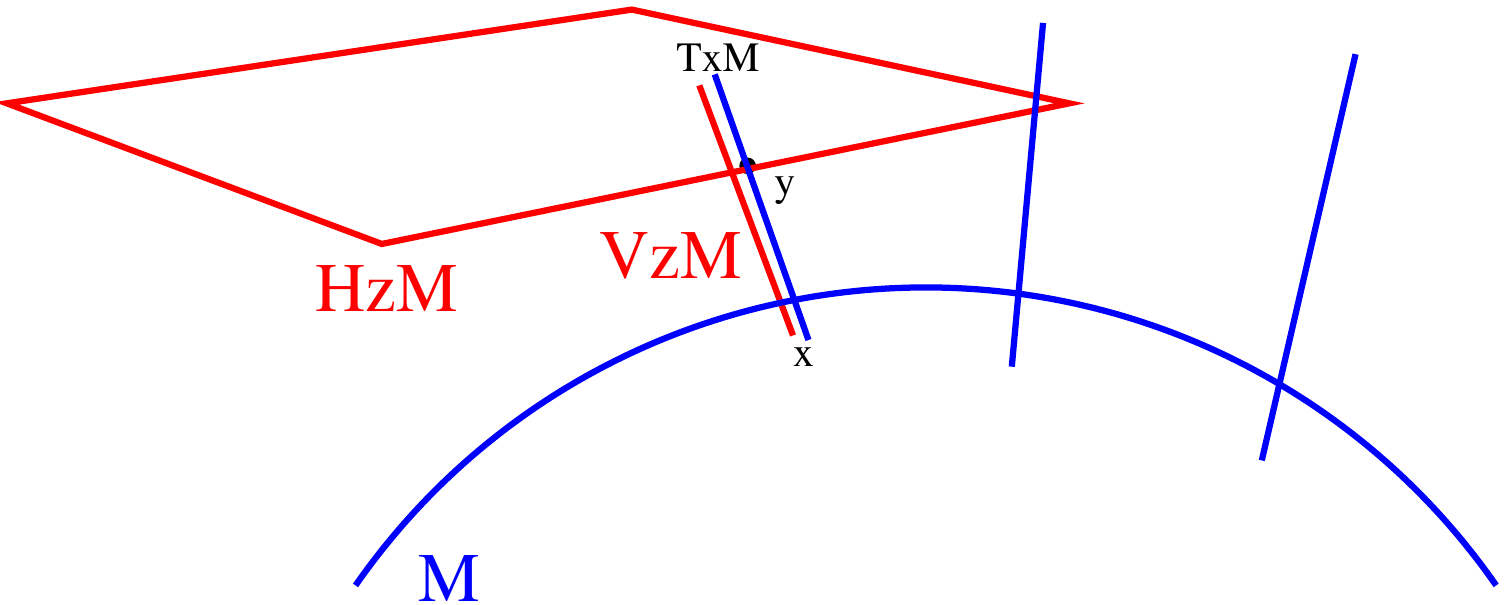}
  \end{center}
  \caption{\label{fig:hv}Horizontal and vertical tangent space to the tangent bundle.}
\end{figure}

The nonlinear connection coefficients split TTM and T*TM into horizontal and vertical space by $\{\delta_a=\partial_a-N^b{}_{a}\bar\partial_b,\bar\partial_a\}$ and $\{dx^a,\delta y^a=dy^a+N^a{}_{v}dx^b\}$, as displayed in figure \ref{fig:hv}. The horizontal (co-)tangent space is identified with the (co-)tangent space along the manifold directions.

Timelike observers move on worldlines $x(\tau)\in M$ with trajectory $(x, \dot x)\in TM$ and $\dot x$ in the cone of timelike vectors. A horizontal orthonormal frame defines their time and space directions along the manifold $\{E_a\}=\{E_0=\dot x^a\delta_a,E_\alpha\}; g^F_{(x,\dot x)}(E_\mu,E_\nu)=-\eta_{\mu\nu}$. Measurable quantities are components of horizontal tensors evaluated in this frame at the observers $TM$ position. 

The geometry of Finsler spacetimes is built from tensors on $TM$; hence physical fields coupling to this geometry will be of the same kind. Lagrange densities on $TM$ require the canonical Sasaki-type $TM$-metric $G=-g^F{}_{ab}(dx^adx^b+F^{-2}\delta y^a\delta y^b)$, which allows us to couple field theories to Finsler spacetime geometry as follows: Choose an action for a $p$-form $\phi(x)$ on $(M,g): S[\phi, g]=\int_M\sqrt{g}\mathcal{L}(g,\phi,d\phi)$, use the Lagrangian for a zero homogenous $p$-form field $\Phi(x,y)$ on $(TM,G)$, introduce Lagrange multipliers to restrict the $p$-form field to be horizontal, integrate over the unit tangent bundle $\Sigma=\{(x,y)\in TM | F(x,y)=1\}$ to obtain the $p$-form field action $S_m[\Phi, L, \lambda]=\int_\Sigma(\sqrt{g^Fh^F} \mathcal{L}(G,\Phi,d\Phi)+\lambda(1-P^H)\Phi)_{|\Sigma}$. Variation yields the  equations of motion, the vanishing of all non horizontal components on shell and the source term of the gravitational dynamics $T_{|\Sigma}$. Our coupling principle ensures that in case the Finsler spacetime is metric, field theories and gravitational dynamics equal those of general relativity.

\section{Gravity}
The geodesic deviation on Finsler spacetimes gives rise to a tensor causing relative gravitational acceleration $\nabla_{\dot x}\nabla_{\dot x}V^a=R^a{}_{bc}(x,\dot x)\dot x^bV^c$. This non-linear curvature given by $R^a{}_{bc}=\delta_{[b}N^a{}_{c]}$ leads to the curvature scalar $\mathcal{R^F}=R^a{}_{ab}y^b$. No further dependence on $L$ or its derivatives appear, thus we choose $\mathcal{R^F}$ as Lagrangian for our Finsler gravity action $S[L,\Phi]=\int_\Sigma(\sqrt{g^Fh^F}\mathcal{R^F})_{|\Sigma}+S_m[L,\Phi]$. Variation with respect to the $L$ yields the Finsler gravity field equation 
\begin{equation}
g^{Fab}\bar\partial_a\bar\partial_bR^F-\frac{6}{F^2}R^F+2g^F_{ab}\big(\nabla_aS_b+S_aS_b+\bar\partial_a(y^q\nabla_q S_b)\big)=-\kappa T_{|\Sigma}\,.
\end{equation}
It contains the curvature scalar, a measure of the departure from metric geometry S, and a Finsler version of the Levi-Civita derivative. In case the function L is the metric length measure the Finsler gravity equation is equivalent to the Einstein equations.

\section{Conclusion}
We constructed a theory of gravity for spacetimes equipped with a general Finsler length measure. In case the Finsler length equals the metric length our theory becomes general relativity, hence all solutions of the Einstein equations are solutions to our Finsler gravity equation. The implications of Finsler spacetime gravity on the dark universe can be studied by spherical symmetric and cosmological solutions that go beyond metric geometry. A perturbative first order Finsler solution around the Schwarzschild and  Friedmann-Robertson-Walker metric is work in progress.

\section*{References}
\bibliography{ProceedingsCPfeifer}

\end{document}